\documentclass[11pt,a4paper]{article}
\usepackage{jcappub}
\usepackage{natbib}

\newcommand{\be}{\,\begin{equation}}
\newcommand{\ee}{\,\end{equation}}

\title{Propagation of galactic cosmic rays in the presence of self-generated turbulence}

\author{Roberto Aloisio$^{1,2}$ and Pasquale Blasi$^{1,2}$}

\affiliation{$^{1}$INAF/Osservatorio Astrofisico di Arcetri, Largo E. Fermi, 5 - 50125 Firenze, Italy\\
$^{2}$Gran Sasso Science Institute (INFN), Viale F. Crispi 7 - 67100 L'Aquila, Italy
}

\emailAdd{aloisio@arcetri.astro.it,blasi@arcetri.astro.it}

\abstract{
Cosmic rays propagating in the Galaxy may excite a streaming instability when their motion is super-alfvenic, thereby producing the conditions for their own diffusion. In this paper we present the results of a self-consistent solution of the transport equation where diffusion occurs because of the self-generated turbulence together with a pre-existing turbulence injected, for instance, by supernova explosions and cascading to smaller scales. All chemicals are included in our calculations, so that we are able to show the secondary to primary ratios in addition to the spectra of the individual elements. All predictions appear to be in good agreement with observations. The fact that data are explained with no need for artificial breaks in the injection spectrum and/or in the diffusion coefficient as functions of momentum can be interpreted as a strong indication that the phenomenon proposed here is in fact being observed in Nature. We also show that there is very good agreement between the calculated proton spectrum and the cosmic ray spectrum inferred from observations of the gamma ray emission from clouds in the Gould's belt. 
} 

\begin{document}
\maketitle

\section{Introduction}
\label{sec:intro}

After decades of investigation, many aspects of the origin of cosmic rays (CRs) remain unclear. Some of the open problems are related to basic principles, while others are more phenomenological in nature and sometimes it is hard to discriminate between the two. While the motion of CRs in the Galaxy is well described as diffusive, and most observables are qualitatively well described within this approach, the fine structure of the theory is all but well defined. The spatial spread of radio and gamma ray emission above and below the disc of the Galaxy clearly suggests that particles have to diffuse perpendicular to magnetic field lines, and/or that magnetic lines must exhibit substantial random walk, as first suggested in Ref. \cite{Jokipii:1969p2496}. Perpendicular diffusion is also needed in order to be consistent with the typical grammage derived from direct measurements of the Boron to Carbon (B/C) ratio and other secondary to primary ratios, since the gas density in the halo is lower than in the disc. The problem of compound diffusion parallel and perpendicular to the background field is not completely solved yet: on one hand, the perpendicular diffusion is determined by the scattering properties in the parallel direction \cite{Bieber:2004p2754}. On the other hand, the statistical properties of the magnetic field (power spectrum, in general different in different directions) are crucial in order to determine the diffusive properties of particles. In addition, all analytical and semi-analytical approaches are limited to the quasi-linear regime, $\delta B/B\ll 1$. Realistic situations in which $\delta B/B\sim 1$ can in general be treated only numerically, but such approaches are usually limited to very high particle rigidities (see for instance \cite{DeMarco:2007p1952}). In addition it has been proposed that cascading of Alfvenic turbulence from large to small scales, needed if to use turbulence for particle scattering, proceeds in a very anisotropic manner: on scales much smaller than the injection scale, most power is in the plane perpendicular to the magnetic field \cite{Goldreich:1995p3018}, a situation that is not very promising for particle scattering. On the other hand, some power may be left in the parallel direction, either as a result of the cascade or as a result of self-generation by CRs. The issue of CR scattering in such turbulence is still being debated (see \cite{Yan:2011p2689,Shalchi:2010p2721}).
 
The overwhelming complexity of this picture has stimulated phenomenological approaches to the problem of CR transport, in which a diffusion coefficient (the same in all directions) is introduced, motivated by the expectation that for $\delta B/B\gg 1$ one would have $D_{\parallel}\sim D_{\perp}$. Numerical approaches do not support such a simple assumption \cite{DeMarco:2007p1952}: even for $\delta B/B\sim 1$, it is found that $D_{\perp}<D_{\parallel}$. Nevertheless we will accept the simple assumption of $D_{\parallel}\sim D_{\perp}\sim D(E)$ as a working hypothesis, though keeping in mind that a better description of CR transport should be sought after. 

A very important piece of the debate on CR diffusion concerns the role of CRs in generating their own scattering centers. This makes the problem of diffusion intrinsically non-linear as was recognized long time ago (e.g. see \cite{Skilling:1975p2176,Holmes:1975p621}). The balance of CR induced streaming instability and damping of the self-generated waves leads to conclude that CRs can be confined in the Galaxy by their own turbulence only for energies below a few hundred GeV. Hence, this phenomenon has not received much attention in recent times.

In addition to these problems of principle, related to the basic nature of magnetic turbulence and the interplay between CRs and magnetic fields, there are phenomenological difficulties, raised as a consequence of more and better data. For instance, recent data from the PAMELA and CREAM experiments \cite{Adriani:2011p10636,Yoon:2011p10500} provide evidence for a change of slope of the spectra of protons and helium nuclei at rigidity $\sim 200$ GV. The spectrum of protons flattens from $E^{-2.85\pm 0.015}$ (for $E< 230$ GeV) to $E^{-2.67\pm 0.03}$ (for $E> 230$ GeV). Although this feature might result from some poorly known aspects of acceleration in the sources, see \cite{Biermann:2010p11467,Ptuskin:2013p11476} for specific models, or from the contribution of different source populations \cite{Zatsepin:2006p11506,Yuan:2011p11541,Thoudam:2012p11567}, in our opinion the most plausible explanation is that something peculiar is happening in terms of propagation of CRs, either in the form of a spatially dependent diffusion \cite{Tomassetti:2012p2445} or because of a competition between different processes relevant for the evolution of waves \cite{Blasi:2012p2344} (or a combination of the last two possibilities).

Despite these difficulties at all levels, propagation of CRs is usually described by using simple diffusion models implemented in propagation codes such as GALPROP \cite{Strong:2007p930} and DRAGON \cite{diBernardo:2010p10666,Evoli:2008p10682}. In the former, diffusion is isotropic and homogeneous (space independent) and peculiar observational features (for instance changes of slope) are usually modeled by assuming artificial breaks in either the injection spectrum of CRs and/or in the diffusion coefficient. In the latter, some effort has been put in introducing a space dependent diffusion coefficient, but in all other respects it has the same advantages and problematic aspects of the former. Neither approach has currently provided a physical explanation of the recent PAMELA and CREAM data. 

Even independent of these complications, the descriptions of CR transport in the Galaxy have always suffered from what is called the anisotropy problem: the B/C ratio which is sensitive to the propagation model can be equally well explained by one of the following models: 1) {\it Standard Diffusion Model} (SDM), in which the injection spectrum is a power law in rigidity $R$ and the diffusion coefficient (or rather the grammage) is a power law $X(R)\propto \beta R^{-\delta}$ (being $\beta$ the particle velocity) with $\delta\simeq 0.6$ for $R>4$ GV and $\delta=0$ for $R<4$ GV. 2) {\it Reacceleration Model} (RAM), in which CRs suffer second order Fermi reacceleration (typically only important for $R<1-10$ GV) and the injection spectrum is a broken power law $Q(R)\propto R^{-2.4}/\left[1 + (R/2)^{2}\right]^{1/2}$ . In both models a break in either the grammage or the injection spectrum is required to fit the data \cite{Ptuskin:2012p2753,Jones:2001p1956}. In the SDM the strong energy dependence of the diffusion coefficient leads to exceedingly large anisotropy for CR energies $\gtrsim$ TeV \cite{Blasi:2012p2053}. In the RAM the anisotropy problem is mitigated \cite{Blasi:2012p2053} although in both models the detailed shape of the anisotropy amplitude as a function of energy is hardly predictable since it is dramatically dependent upon the position of the few most recent and closest sources \cite{Blasi:2012p2053,Lee:1979p1621,Ptuskin:2006p620}. The injection spectrum required at high energy by the SDM is roughly compatible with the one expected based on diffusive shock acceleration (DSA), while the one required by the RAM is too steep and requires non-standard versions of the acceleration theory, possibly invoking the role of the finite velocity of scattering centers (see \cite{Caprioli:2010p133,Ptuskin:2010p1025,Caprioli:2012p2411} for a discussion of this issue).

In the present paper we follow up on previous work \cite{Blasi:2012p2344} in which it has been proposed that the changes of slope in the spectra of protons observed by PAMELA and CREAM may reflect the interplay between self-generation of turbulence by CRs and turbulent cascade from large scales. The transition from diffusion in the self-generated turbulence to diffusion in pre-existing turbulence naturally reflects in a spectral break for protons from steep (slope $\sim 2.9$) to flatter (slope $\sim 2.65$) at rigidity $R\sim 200$ GV \cite{Blasi:2012p2344}. This break compares well with the protons data as collected by the PAMELA satellite \cite{Adriani:2011p10636}, and the high energy slope is in agreement with the one measured by the CREAM balloon experiment \cite{Yoon:2011p10500}. At rigidity $R<10$ GV the spectra show a hardening reflecting the advection of particles with Alfv\'en waves \cite{Blasi:2012p2344}.

A previous attempt at taking into account the self-generated turbulence around the sources of Galactic CRs was made by \cite{Ptuskin:2008p2148}, but the effect of pre-existing turbulence was not considered. In \cite{Ptuskin:2006p2981} the authors investigate the possibility that a Kraichnan-type cascade may be suppressed at $k\gtrsim 10^{-13} cm^{-1}$ as a result of resonant absorption of the waves by CRs. The authors claim that this effect may account for the low energy behaviour of the B/C ratio. 

Since the problem of CR diffusion that we intend to investigate is intrinsically non-linear, it is obvious that many of the oversimplifications discussed above need to be assumed here as well. However we try to make one step towards a physical understanding of the nature of CR diffusion in the Galaxy and of the role of CRs in their own transport. All nuclei from hydrogen to iron (including all stable isotopes) and their spallation and ionization losses are included in the calculations, so that we can compare our results with existing B/C data, as well as with the observed spectra of the individual elements. We retain the assumption that diffusion may be approximated as homogeneous and isotropic within the propagation model. 

The paper is structured as follows: in \S \ref{sec:self} we describe the process of self-generation of turbulence induced by CRs and its interaction with the turbulent cascade from larger scales. In \S \ref{sec:slab} we describe a generalization of the weighted slab model of Ref. \cite{Jones:2001p1956} to the case in which the diffusion coefficient is non-linearly generated as discussed in \S \ref{sec:self}. In this model the spectra, chemical composition and grammage traversed by particles are functions of each other. Our results are illustrated in \S \ref{sec:results}. In \S \ref{sec:clouds} we compare our CR spectrum with the one derived in  \cite{Neronov:2012p2217,Kachelrie:2012p10950} by using gamma ray data from clouds in the Gould belt. The implications of the non linear CR transport proposed here are discussed in \S \ref{sec:discuss}.

\section{Equation for waves with self-generation}
\label{sec:self}

Following Ref. \cite{Blasi:2012p2344} we consider waves responsible for CR scattering as originated through two processes: 1) turbulent cascade of power injected by supernova remnants (SNRs) at large scales, and 2) self-generated waves produced by CRs through streaming instability. The process of cascading is all but trivial to model, in that it is known that power is channeled in a complex way into parallel and perpendicular wave numbers. It is not the purpose of this paper to explore these aspects in detail. We rather adopt the view that cascading occurs through non-linear Landau damping (NLLD) as modeled in \cite{Zhou:1990p2226}, namely through diffusion in $k$-space with a diffusion coefficient which depends on the wave spectrum $W(k)$ (this reflects the intrinsic non-linearity of the phenomenon). The diffusion coefficient in $k$-space can be written as \cite{Zhou:1990p2226,Miller:1995p3113}:
\be
D_{kk} = C_{\rm K} v_{\rm A} k^{\alpha_1} W(k)^{\alpha_2},
\label{eq:dkk}
\ee
where $C_{\rm K}\approx 5.2\times 10^{-2}$ \cite{Ptuskin:2003p2174} and $v_{\rm A}=B_{0}/\sqrt{4\pi \rho_{ion}}$ is the Alfv\'en speed in the unperturbed magnetic field (notice that $\rho_{ion}$ is the density of the ionized gas in the Galaxy, not the total gas density). In the present paper we assume a Kolmogorov phenomenology for the cascading turbulence, so that $\alpha_1=7/2$ and $\alpha_2=1/2$, and an unperturbed magnetic field $B_0=1$ $\mu$G. The value of the magnetic field used in this paper has to be interpreted as the mean value calculated on the volume of the Galactic halo. This is clearly smaller than the magnetic field measured in the disc of the Galaxy. It is interesting to note that, because of the assumption of homogeneous diffusion, the limitations derived from assuming a mean field throughout the Galaxy are in fact also common to more sophisticated propagation calculations, such as GALPROP \cite{Strong:2007p930} and DRAGON \cite{diBernardo:2010p10666,Evoli:2008p10682}. 

The equation describing wave transport along the direction of the ordered magnetic field can be written as follows\footnote{If the wave spectrum $W$ were the omnidirectional spectrum (in 3 dimensions) the damping term would read $\frac{\partial}{\partial k}\left[k^{2} D_{kk} \frac{\partial}{\partial k} (W/k^{2}) \right]$. Here however we assume that all waves move along the direction of the background magnetic field.} \cite{Miller:1995p3113}:
\be
\frac{\partial}{\partial k}\left[ D_{kk} \frac{\partial W}{\partial k}\right] + \Gamma_{\rm CR}W = q_{W}(k), 
\label{eq:cascade}
\ee
where $q_{W}(k)$ is the injection term of waves with wavenumber $k$. In principle this equation should also contain advection terms of the type $v_{A}\partial W/\partial z$, which we neglected here, for several reasons: 1) the advection time scale is of order $H/v_{A}$ which for typical values of the parameters is much longer than any other time scale in our problem (see discussion in \S \ref{sec:spectra}); 2) the simplest solution of the transport equation for cosmic rays in the Galaxy has a space derivative which is independent of $z$ so that the growth rate of waves, described by Eq. \ref{eq:gammacr}, is also independent of z. Moreover the damping term is also assumed to be independent of $z$, hence $W$ is expected to be constant in $z$ within such simple framework. In the present calculations we assume that waves are only injected on a scale $l_{c}\sim 50-100$ pc, for instance by supernova explosions. This means that $q_{W}(k)\propto \delta (k-1/l_{c})$. Notice that the wave injection is also assumed to occur uniformly in the whole halo of the Galaxy. The level of pre-existing turbulence is normalized to the total power $\eta_B=\delta B^{2}/B_0^2 = \int dk W(k)$. Strictly speaking the wave number that appears in this formalism is the one in the direction parallel to that of the ordered magnetic field. In a more realistic situation in which most power is on large spatial scales, the role of the ordered field is probably played by the local magnetic field on the largest scale.

The term $\Gamma_{\rm CR}W$ in Eq. (\ref{eq:cascade}) describes the generation of wave power through CR induced streaming instability, with a growth rate \cite{Skilling:1975p2176}:
\be
\Gamma_{\rm cr}(k)=\frac{16 \pi^{2}}{3} \frac{v_{\rm A}}{k\,W(k) B_{0}^{2}} \sum_{\alpha} \left[ p^{4} v(p) \frac{\partial f_{\alpha}}{\partial z}\right]_{p=Z_{\alpha} e B_{0}/kc} ,
\label{eq:gammacr}
\ee 
where $\alpha$ is the index labeling nuclei of different types. All nuclei, including all stable isotopes for a given value of charge, are included in the calculations. As discussed in much previous literature, this is very important if to obtain a good fit to the spectra and primary to secondary ratios, especially B/C. The growth rate, written as in Eq. \ref{eq:gammacr}, refers to waves with wave number $k$ along the ordered magnetic field. It is basically impossible to generalize the growth rate to a more realistic field geometry by operating in the context of quasi-linear theory, therefore we will use here this expression but keeping in mind its limitations.

The waves produced by the excitation of the streaming instability move along the direction of a decreasing CR density. In the context of this formalism this direction coincides with the $z$-direction, and there are no waves moving in the opposite direction. If there were waves moving in the opposite direction, the growth rate Eq. \ref{eq:gammacr} would contain a second term, proportional to the energy density of waves moving in the direction of growing CR density.

Streaming instability proceeds in a resonant manner, therefore the production of waves with a given wavenumber $k$ is associated with particles with momentum such that their Larmor radius equals $1/k$.  Non-resonant modes have also been discussed in the literature (for instance see Ref. \cite{Bell:2004p737}) but their growth rate is negligible in the context of CR propagation in the Galaxy. 

A wave with wavenumber $k$ can be either produced or absorbed resonantly by nuclei that satisfy the resonance condition. Hence it is clear that the diffusion coefficient relevant for a given nucleus is the result of the action of all others (see Eq. (\ref{eq:gammacr})). In practice, since protons and helium are the most abundant species, the diffusion coefficients for protons and helium are strongly affected by the presence of both elements. To first approximation, the diffusion coefficient for heavier nuclei is mainly determined by waves produced by protons and helium nuclei. The contribution of all other stable isotopes has a sizable effect, at the level of few percent.

The solution of Eq. (\ref{eq:cascade}) can be written in an implicit form 
\be
W(k) = \left [ W_0^{1+\alpha_2}\left (\frac{k}{k_0}\right )^{1-\alpha_1} + \frac{1+\alpha_2}{C_{\rm K}v_A}\int_k^\infty\frac{dk'}{k'^{\alpha_2}}\int_{k_0}^{k'} dk'' \Gamma_{CR}(k'') W(k'') \right ]^{\frac{1}{1+\alpha_2}} ,
\label{eq:waves}
\ee
being $k_0=1/l_c$. The two terms in Eq. (\ref{eq:waves}) refer respectively to the pre-existing magnetic turbulence and the CR induced turbulence. In the limit in which there are no CRs (or CRs do not play an appreciable role) one finds the standard Kolmogorov wave spectrum 
\be
W(k)=W_0\left (\frac{k}{k_0} \right)^{-s} ~~~ s=\frac{\alpha_1-1}{\alpha_2+1} = \frac{5}{3}
\ee
normalized, as discussed above, to the total power $W_0=(s-1)l_c\eta_B$.

The diffusion coefficient relevant for a nucleus $\alpha$ can be written as:
\be
D_{\alpha} (p) = \frac{1}{3} \frac{pc}{Z_\alpha eB_{0}} v(p) \left[ \frac{1}{k\ W(k)} \right]_{k=Z_{\alpha} e B_{0}/pc},
\label{eq:diff}
\ee
where $W(k)$ is the power spectrum of waves given by Eq. (\ref{eq:waves}). The non-linearity of the problem is evident here: the diffusion coefficient for each nuclear specie depends on all other nuclei through the wave power $W(k)$, but the spectra are in turn determined by the relevant diffusion coefficient. The problem can be closed, at least in an implicit way, by writing the transport equation for each nucleus, which will be done in the next section. 

\section{The modified weighted slab model with self-generated diffusion}
\label{sec:slab}

The transport equation for nuclei of type $\alpha$ can be written as:
\be
-\frac{\partial}{\partial z} \left[D_{\alpha}(p) \frac{\partial f_{\alpha}}{\partial z}\right] + v_{A} \frac{\partial f_{\alpha}}{\partial z}-\frac{dv_{A}}{dz}\frac{p}{3}\frac{\partial f_{\alpha}}{\partial p}
+\frac{f_{\alpha}}{\tau_{sp,\alpha}} + \frac{1}{p^{2}} \frac{\partial}{\partial p}\left[ p^{2} \left(\frac{dp}{dt}\right)_{\alpha,ion} f_{\alpha}\right] =\\ q_{\alpha}(p,z)+\sum_{\alpha'>\alpha} \frac{f_{\alpha'}}{\tau_{sp,\alpha'}},
\label{eq:transport}
\ee
where $\tau_{\alpha,sp}$ is the time scale for spallation of nuclei of type $\alpha$, $q_{\alpha}(p,z)$ is the rate of injection per unit volume of nuclei of type $\alpha$ and $\left(\frac{dp}{dt}\right)_{\alpha,ion}$ is the corresponding rate of ionization losses. Following Ref. \cite{Jones:2001p1956} we introduce a surface gas density in the disc $\mu=2h_{d} n_{d} m$, where $h_{d}$ is the half-thickness of the Galactic disc and $n_{d}$ is the gas density in the disc, in the form of particles with mean mass $m$. The measured value of the surface density is $\mu\approx 2.4$  mg/cm$^{2}$ \cite{Ferriere:2001p10947}, that corresponds to a gas density $n_d\approx 1$ cm$^{-3}$ assuming a galactic disc half thickness of $150$ pc, and a chemical composition of the ISM made of 85\% of hydrogen and 15\% helium. 

The transport equation in Eq. \ref{eq:transport} is based on the assumption, described above, that all waves are generated along $z$ and in the direction away from the Galactic disc (if there were an equal number of waves moving toward the disc, then all terms proportional to $v_{A}$ would vanish). This assumption is formally justified only for the self-generated waves, while the one deriving from wave-wave coupling can move in all directions, therefore when this process dominates in principle one should expect that the advection term disappears. However, as we show below, the transition from self-generated waves to pre-exisiting turbulence occurs at several hundred GV, where advection plays no role in any case. Therefore we preferred, for the sake of simplicity, to keep the equation in the same form at high energy as it has in the low energy regime, where advection is important. Given the random orientation of the magnetic field on large scales, one should also keep in mind that the effective advection velocity of self-generated waves in the $z$ direction could be somewhat smaller than $v_{A}$.

It is worth stressing that the CR advection with the waves (important at low energies) is not in contradiction with the fact that we neglected the advection term in the wave equation Eq. \ref{eq:cascade}, as discussed in detail \S \ref{sec:spectra}: in the energy region below a few tens GeV the diffusion time scale becomes comparable with $H/v_{A}$ despite the fact that wave growth and damping occur on much smaller scales.

With this formalism, and assuming an infinitely thin disc, we can write the transport equation as:
$$
-\frac{\partial}{\partial z} \left[D_{\alpha}(p) \frac{\partial f_{\alpha}}{\partial z}\right] + v_{A} \frac{\partial f_{\alpha}}{\partial z}-\frac{2}{3}v_{A}\delta(z) p \frac{\partial f_{\alpha}}{\partial p}
+\frac{\mu v(p) \sigma_{\alpha}}{m}\delta(z) f_{\alpha} + \frac{1}{p^{2}} \frac{\partial}{\partial p}\left[ p^{2} \left(\frac{dp}{dt}\right)_{\alpha,ion} f_{\alpha}\right] =
$$
\be
= 2 h_d q_{0,\alpha}(p) \delta(z) +\sum_{\alpha'>\alpha} \frac{\mu v(p) \sigma_{\alpha'\to\alpha}}{m}\delta(z) f_{\alpha'},
\label{eq:slab}
\ee
where $\sigma_{\alpha}$ is the spallation cross section of a nucleus of type $\alpha$ and $q_{0,\alpha}(p)$ is the rate of injection per unit volume in the disc of the Galaxy. The total cross section for spallation and the cross sections for the individual channels of spallation of a heavier element to a lighter element ($\sigma_{\alpha'\to\alpha}$) have been taken from Refs. \cite{Webber:1990p3045,Webber:2003p3044}. As stressed above, for the sake of a meaningful comparison with data, it is important to take into account the stable isotopes of all elements. This is important not only for pure secondary elements, namely elements produced only through spallation, such as Boron (B=$^{10}$B+$^{11}$B), but also for Nitrogen (N=$^{14}$N+$^{15}$N), which gets a significant secondary contribution from spallation of heavier nuclei, for Carbon (C=$^{12}$C+$^{13}$C) and Oxigen (O=$^{16}$O+$^{17}$O+$^{18}$O) \cite{diBernardo:2010p10666,Evoli:2008p10682}. Moreover, for rapidly decaying isotopes, i.e. the ones that decay on times much shorter than their escape time from the Galaxy, we assume that the decay is instantaneous. This means that in the sum over $\alpha'$ in the rhs of Eq.(\ref{eq:slab}) we consider also terms of the type $\sigma_{\alpha'\to\alpha''} f_{\alpha'}$ being $\alpha''$ a nuclear specie that rapidly decays into $\alpha$. 

Here $v(p)=\beta c$ is the velocity of nuclei of type $\alpha$ having momentum $p$. Notice that since the gas is assumed to be present only in the disc, and the ionization rate is proportional to the gas density, one can write: $\left(\frac{dp}{dt}\right)_{\alpha,ion}=2h_d \delta(z) b_{0,\alpha}(p)$, where $b_{0,\alpha}(p)$ contains the particle physics aspects of the process (see \cite{Strong:1998p10532} and references therein for a more detailed discussion of this term).  

Integrating Eq. (\ref{eq:slab}) between $z=0^{-}$ and $z=0^{+}$ and recalling that $(\partial f/\partial z)_{0^{-}}=-(\partial f/\partial z)_{0^{+}}$, one obtains:
$$
-2D_{\alpha}(p) \left(\frac{\partial f_{\alpha}}{\partial z}\right)_{z=0^{+}} - 
\frac{2}{3}v_{A}p \frac{\partial f_{0,\alpha}}{\partial p}
+ \frac{\mu v(p) \sigma_{\alpha}}{m} f_{0,\alpha} + \frac{2 h_d}{p^{2}} \frac{\partial}{\partial p}\left[ p^{2} b_{0,\alpha}(p) f_{0,\alpha}\right] = 
$$
\be
= 2 h_d q_{0,\alpha}(p) + \sum_{\alpha'>\alpha} \frac{\mu v(p) \sigma_{\alpha'\to\alpha}}{m}f_{0,\alpha'},
\label{eq:slab1}
\ee
For $|z|>0$ Eq. (\ref{eq:slab}) implies that
\be
\left[v_{A} f_{\alpha}-D_{\alpha}(p) \frac{\partial f_{\alpha}}{\partial z}\right] = Constant \rightarrow f_{\alpha}(z,p)=f_{0,\alpha}(p) 
\frac{1-\exp\left[-\frac{v_{A}}{D_{\alpha}}(H-z)\right]}{1-\exp\left[-\frac{v_{A}}{D_{\alpha}}H\right]},
\ee
where we have used the fact that $f_{\alpha}(z\to 0,p)=f_{0,\alpha}(p)$ and have adopted the free escape boundary condition from a halo of half-height $H$: $f_{\alpha}(z=\pm H,p)=0$. 

Following Ref. \cite{Jones:2001p1956} we introduce the function $I_{\alpha}(E)$ (flux of nuclei with kinetic energy per nucleon $E$ for nuclei of type $\alpha$), such that $I_{\alpha}(E)dE=v p^{2} f_{0,\alpha}(p)dp$. It is easy to show that $I_{\alpha}(E)=A_{\alpha}p^{2} f_{0,\alpha}(p)$, being $A_\alpha$ the atomic mass number of the nucleus.

Substituting this finding in Eq. (\ref{eq:slab1}) we easily obtain:
$$
\frac{I_{\alpha}(E)}{X_{\alpha}(E)} + \frac{d}{dE}\left\{\left[ \left(\frac{dE}{dx}\right)_{ad} +  \left(\frac{dE}{dx}\right)_{ion,\alpha}\right] I_{\alpha}(E)\right\} + \frac{\sigma_{\alpha} I_{\alpha}(E)}{m} =
$$
\be
= 2 h_d \frac{A_{\alpha} p^{2} q_{0,\alpha}(p)}{\mu v} + \sum_{\alpha'>\alpha} \frac{I_{\alpha}(E)}{m}\sigma_{\alpha'\to\alpha},
\label{eq:slab2}
\ee
where 
\be
X_{\alpha}(E) = \frac{\mu v}{2 v_{A}} \left[ 1-\exp\left(-\frac{v_{A}}{D_{\alpha}}H\right)\right]
\ee
is the grammage for nuclei of type $\alpha$ with kinetic energy per nucleon $E$ and 
\be
\left(\frac{dE}{dx}\right)_{ad} = -\frac{2 v_A}{3\mu c} \sqrt{E(E+m_p c^2)} 
\ee
is the rate of adiabatic energy losses due to advection.

Eq. (\ref{eq:slab2}) can be transformed in the form $\Lambda_{1,\alpha}(E)I_{\alpha}(E)+\Lambda_{2,\alpha}(E)\partial_E I_{\alpha}(E) = Q_{\alpha}(E)$, with obvious meaning of the $\Lambda_{i,\alpha}(E)$, that, assuming the boundary condition $I(E\to\infty)=0$, has the solution
\be
I_{\alpha}(E)=\int_{E} dE' \frac{Q_{\alpha}(E')}{\left |\Lambda_{2,\alpha}(E')\right |}\exp{\left [-\int_E^{E'} dE'' \frac{\Lambda_{1,\alpha}(E'')}{\left |\Lambda_{2,\alpha}(E'')\right |}\right ]}~.
\ee 
The injection term $Q_\alpha(E)$ can be written assuming a simple injection model in which all CRs are produced by SNRs with the same power law spectrum:  
\be
Q_{\alpha}(E)=\frac{A_{\alpha} p^2}{\mu v} \frac{\xi_{\alpha}E_{SN}{\cal R}_{SN}}{\pi R_d^2 \Gamma(\gamma) c (m_p c)^4}  \left (\frac{p}{m_p c}\right )^{-\gamma}+ \sum_{\alpha'>\alpha} \frac{I_{\alpha}(E)}{m}\sigma_{\alpha'\to\alpha},
\ee
where $\xi_{\alpha}$ is the fraction of the total kinetic energy of a supernova ($E_{SN}=10^{51}$ erg) channelled into CRs of specie $\alpha$, $R_d=10$ kpc is the radius of the Galactic disk, ${\cal R}_{SN}=1/30$ y$^{-1}$ is the rate of SN explosions and $\Gamma(\gamma)=4\pi\int_{0}^{\infty}  dx x^{2-\gamma}[\sqrt{x^2+1} -1 ]$ comes from the normalization of the CR spectrum. 

The equations for the waves and for CR transport are solved together in an iterative way, so as to return the spectra of particles and the diffusion coefficient for each nuclear specie and the associated grammage. The procedure is started by choosing guess injection factors for each type of nuclei, and a guess for the diffusion coefficient, which is assumed to coincide with the one predicted by quasi-linear theory in the presence of a background turbulence. The first iteration returns the spectra of each nuclear specie and a spectrum of waves, that can be used now to calculate the diffusion coefficient self-consistently. The procedure is repeated until convergence, which is typically reached in a few steps, and the resulting fluxes and ratios are compared with available data. This allows us to renormalize the injection rates and restart the whole procedure, which is repeated until a satisfactory fit is achieved. Since the fluxes of individual nuclei affect the grammage through the rate of excitation of streaming instability and viceversa the grammage affects the fluxes, the procedure is all but trivial. 

\section{Results}
\label{sec:results}

In this section we illustrate the main results of our calculations in terms of spectra, grammage and secondary/primary and primary/primary ratios. In \S \ref{sec:spectra} we show the spectra of protons, helium and heavier primary nuclei. In the non-linear calculations carried out here, the CR spectra determine, and are in turn determined, by the diffusion coefficient and grammage. The diffusion coefficient and the grammage traversed by CRs is discussed in \S \ref{sec:grammage}. An important check of the self-consistency of the calculations is represented by the ratio of fluxes of secondary and primary nuclei, as illustrated in \S \ref{sec:ratios}. Finally in \S \ref{sec:clouds} we discuss the comparison of our predicted proton spectrum with the one derived from observations of the gamma ray emission from clouds in the Gould's belt. All plots refer to a benchmark case with a halo size $H=4$ kpc.

\subsection{Spectra of primary nuclei}
\label{sec:spectra}

\begin{figure}
   \centering
   \includegraphics[width=0.6\textwidth]{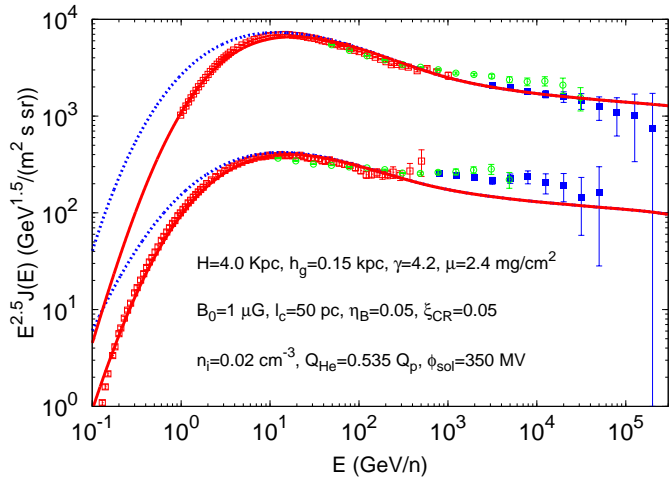}
   \caption{Spectra of protons and helium nuclei for the values of parameters as indicated. The solid (red) lines show the spectra at the Earth, while the dotted (blue) lines show the spectra in the ISM, namely before correction for solar modulation. Data points are from PAMELA (open squares) \cite{Adriani:2011p10636}, CREAM (filled squares) \cite{Yoon:2011p10500} and ATIC-2 (open circles) \cite{Panov:2009p11401}.}
   \label{fig:spectra} 
\end{figure}

The calculated spectra of protons and helium nuclei are plotted in Fig. \ref{fig:spectra}. The dashed (blue) lines are the spectra in the ISM while the continuous (red) lines represent the spectra after correction for solar modulation, modeled as in \cite{diBernardo:2010p10666}. The solar modulation parameter has been taken here to be $\Phi=350$ MV. The predicted spectra are compared with the PAMELA \cite{Adriani:2011p10636}, CREAM \cite{Ahn:2009p10593,Ahn:2010p624,Yoon:2011p10500}, ATIC-2 \cite{Panov:2009p11401} and HEAO data \cite{Engelmann:1990p10596}. The predicted proton spectrum is in excellent agreement with PAMELA and CREAM data at all energies and clearly shows a hardening at energies above $\sim 200$ GeV. The ATIC-2 data appear to show a harder spectrum compared with PAMELA and CREAM. The helium spectrum is also in good agreement with data up to a few hundred GeV/nucleon. At higher energies the measurements are dominated by CREAM and ATIC-2 data and the agreement is poorer. Whether this discrepancy is due to a flatter injection of helium nuclei or to a different systematic error between the PAMELA and the CREAM/ATIC-2 data remains an open question. At low energies we do not see a clear evidence for any harder injection spectrum of helium nuclei compared with protons, as shown by the excellent fit to the PAMELA data. One should recall that a $\sim 20\%$ systematic error in the energy determination in any of these experiments would reflect in a factor $\sim 1.5$ at $\sim $TeV energies in the absolute fluxes as plotted in Fig. \ref{fig:spectra}, due to the multiplication factor $E^{2.5}$.

In Ref. \cite{Blasi:2012p2053} it was suggested that a mild hardening of the He spectrum could be due to spallation. However, as pointed out in \cite{Vladimirov:2012p3043}, this requires a gas density larger than usual and a cross section for spallation larger than in Ref. \cite{Webber:1990p3045,Webber:2003p3044}. 

From the physics point of view one may envision that details of the mechanism of DSA in supernova remnants can possibly explain a slightly flatter injection spectrum of helium nuclei compared with hydrogen (see for instance \cite{Malkov:2012p2391}). Although possible, this explanation would suggest that subtle model dependent aspects of acceleration physics may have macroscopic effects. Whatever the explanation, it is however clear that the problem of the hardening of the proton and helium spectra at $\sim 200$ GV rigidity is separate from that of a possibly systematically harder He spectrum. In fact, the former effect is clearly visible in the spectra of all nuclei. 

In our calculations, this effects stems from the change of transport regime, from diffusion in the self-generated turbulence to diffusion in pre-existing turbulence. The transition translates in a clear change of slope at $\sim 200$ GeV/n in the spectra observed at Earth. In the low energy region, after accounting for solar modulation, the agreement with data is excellent down to sub-GeV energies. At rigidity below $\sim 10$ GV the CR transport becomes dominated by advection with Alfv\'en waves. This is the reason for the flattening of the unmodulated spectra at low energies. 

\begin{figure}
   \centering
   \includegraphics[width=0.6\textwidth]{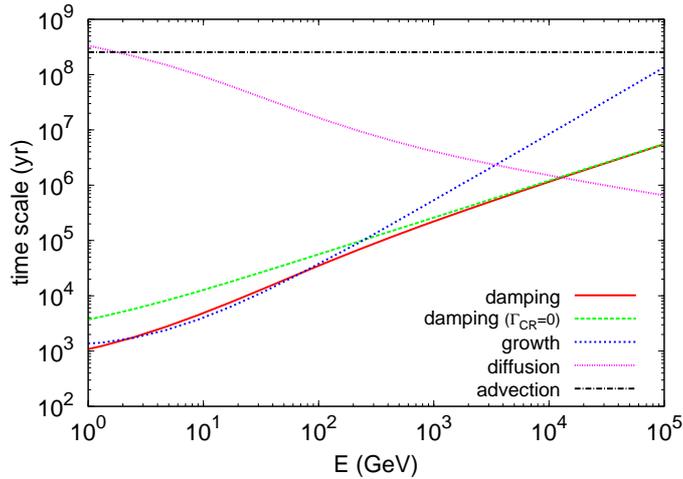}
   \caption{Time scales for: wave damping $\tau_{damp}=k^{2}/D_{kk}$ (with and without the effect of CR), wave growth $\tau_{cr}=1/\Gamma_{cr}$, wave advection $\tau_{adv}=H/v_{A}$ and diffusion $\tau_{diff}=H^{2}/D(E)$ (for protons). On the x-axis we plot the energy of protons resonating with the waves of a corresponding wavenumber $k$, as far as growth and damping are concerned.}
   \label{fig:times} 
\end{figure}

The transition between the propagation in a background of waves affected by CR-induced growth and propagation in waves mainly affected by cascading may be better appreciated by looking at the time scales for wave growth $\tau_{cr}=1/\Gamma_{cr}$, for damping $\tau_{damp}=k^{2}/D_{kk}$, wave advection $\tau_{adv}=H/v_{A}$ and diffusion $\tau_{diff}=H^{2}/D(E)$. We show these time scales in Fig. \ref{fig:times}, where we refer to protons as a reference case. The change of slope of the CR diffusion time with energy at $\sim 200$ GV can easily be seen. The diffusion time becomes longer than the advection time, $H/v_{A}$ below 10 GV, which is the reason for the low energy hardening. Notice that all time scales relevant for wave evolution (growth and damping) are much smaller than $H/v_{A}$ at all energies, which implies that while advection may be important for CR evolution it is unlikely to have any effect on the wave spectrum, so as to justify {\it a posteriori} the fact that we neglect the term $v_{A}\partial W/\partial z$ in Eq. \ref{eq:cascade}. Notice also that all times scales listed above depend on $W(k)$ which in turn is determined by the propagation of all nuclear species, although protons and helium nuclei play the most important role. At high energies, where self-generation of waves has a negligible effect, the damping time in Fig. \ref{fig:times} equals the result that one would obtain with a pure Kolmogorov spectrum for $W(k)$ (green long-dashed line). Finally, the reader should appreciate that in the low energy regime where advection dominates, the distribution function of CRs has a spatial gradient that asymptotically vanishes (for $E \ll 10$ GeV), therefore the growth rate also vanishes in the same regime: CR induced wave growth is present only if CRs drift faster that Alfv\'en waves. This is the reason why the expression for the growth rate that is most commonly found in the literature is $\Gamma_{cr}\propto n_{cr}(p>p_{res})(v_{D}-v_{A})$ for drift velocity $v_{D}>v_{A}$ and $\Gamma_{cr}=0$ for $v_{D}>v_{A}$. The latter condition becomes increasingly more satisfied in realistic calculations when advection of CRs with the waves becomes prominent. In other words, for very low energies (large values of $k$) waves are not generated by CRs but rather produced by the cascading process from smaller $k$'s. The self-generated growth appears to be a potentially very important process for CR propagation only for energies between $\sim 10$ GeV and $\sim 1000$ GeV.

In the range of energies $10\lesssim E\lesssim 200$ GeV/n the self-generation is so effective as to make the diffusion coefficient have a steep energy dependence (see below). As a consequence the injection spectrum that is needed to fit the data is $p^{2}q(p)\propto p^{-2.2}$, which is not far from what can be accounted for in terms of DSA if the velocity of the scattering centers is taken into account \cite{Caprioli:2010p133,Ptuskin:2010p1025,Caprioli:2012p2411}. No break in the injection spectrum is imposed by hand throughout our calculations. 

\begin{figure}
   \centering
   \includegraphics[width=0.6\textwidth]{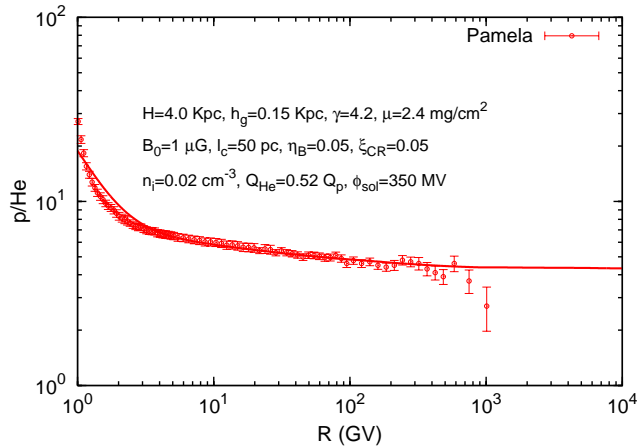}
   \caption{Ratio of fluxes of protons and helium nuclei for the values of parameters as indicated. PAMELA data points are from \cite{Adriani:2011p10636}.}
   \label{fig:p-He-ratio} 
\end{figure}

In Fig. \ref{fig:p-He-ratio} we show the predicted ratio of fluxes of protons and He. The ratio is compared with the one measured by PAMELA. The agreement between the two is very good, with the possible exception of the last two high energy points where however the error bars are rather large. At even higher rigidity, the discrepancy increases, reflecting the fact that our predicted helium flux is lower than the one measured by CREAM in the TeV range.
 
\begin{figure}
   \centering
   \includegraphics[width=1.\textwidth]{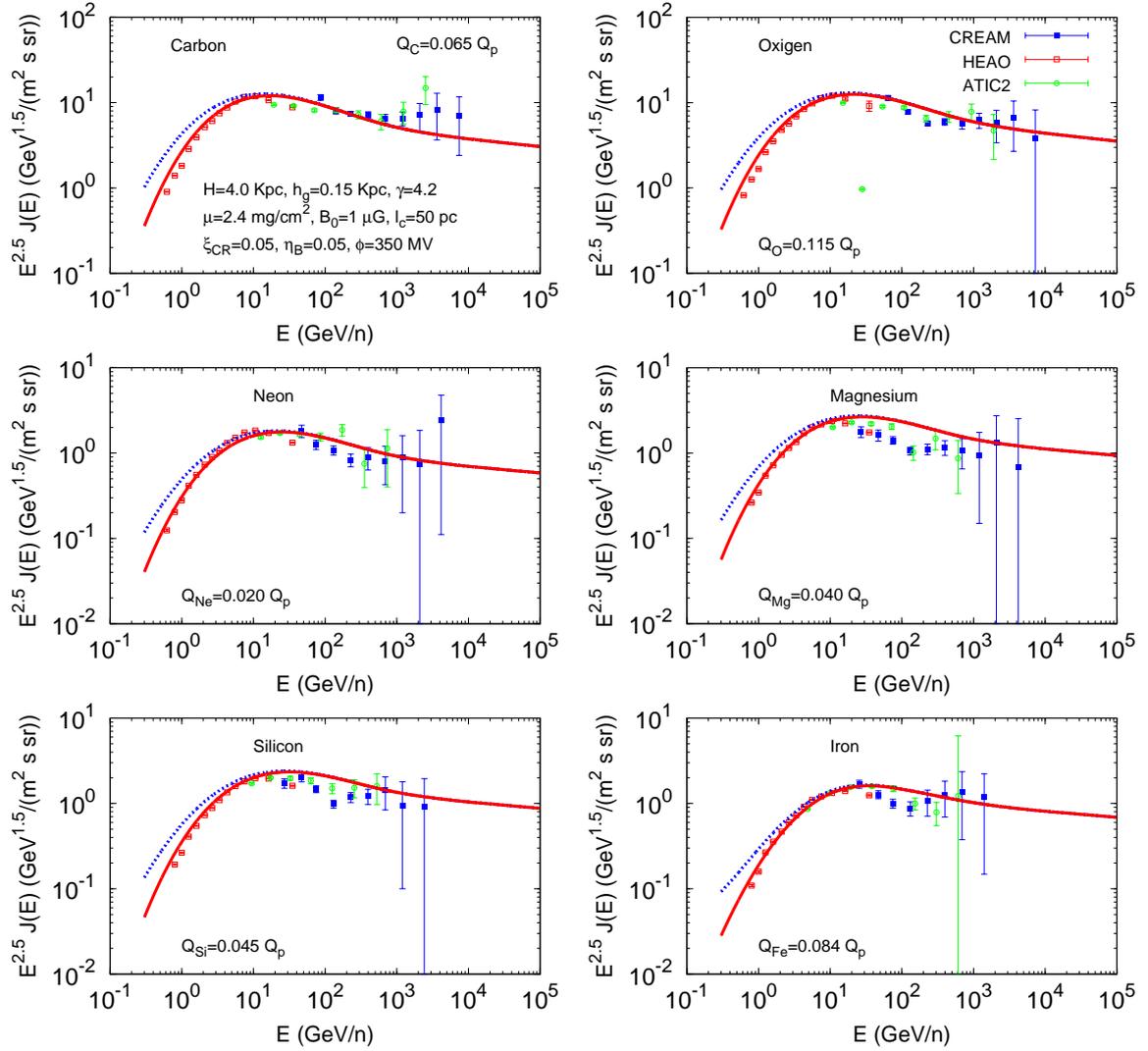}
   \caption{Fluxes of primary nuclei. The solid (red) lines show the spectra at the Earth, while the dotted (blue) lines show the spectra in the ISM, namely before correction for solar modulation. The same solar modulation potential $\Phi$ as for protons and He nuclei has been chosen, despite the fact that these data were collected at different times. Data points are from CREAM \cite{Ahn:2009p10593,Ahn:2010p624}, HEAO \cite{Engelmann:1990p10596} and ATIC-2 \cite{Panov:2009p11401}.}
   \label{fig:nuclei} 
\end{figure}

The spectra of primary nuclei are shown in Fig. \ref{fig:nuclei}. The agreement between the predicted and the observed spectra is reasonably good. All spectra have the same trend, steeper below a few hundred GV and harder at higher rigidity, with a hardening below $\sim 10$ GV, which is again due to the effect of advection with waves, namely the term $v_{A}\partial f_{\alpha}/\partial z$ in the trasport equation. The spectrum of nitrogen is not plotted in Fig. \ref{fig:nuclei}, since it has approximately a half primary and half secondary origin, as discussed below (\S \ref{sec:ratios}).

In all plots in Fig. \ref{fig:nuclei} the solid (red) line refers to the fluxes at the Earth (after correction for solar modulation), while the dotted (blue) lines refer to the spectra in the ISM. The low energy part of the spectra at Earth are obtained by using the same level of solar modulation adopted for protons and helium nuclei (Fig. \ref{fig:spectra}). We made this choice for consistency with the rest of the paper, but one should keep in mind that in principle different data sets were collected at different times and thus at different stages of solar activity. Therefore it could well be that slightly different values of the parameter $\Phi$ should be used for different sets of data, and this would likely improve further the agreement between predicted and observed spectra. 

The CR acceleration efficiency in terms of protons, that is needed to ensure the level of wave excitation necessary to explain observations, is $\xi_{CR}\sim 5\%$, perfectly in line with the standard expectation of the so-called SNR paradigm. The quantities $Q_{\alpha}$ shown in the plots represent the fraction of nuclei of type $\alpha$ injected at the source relative to proton injection $Q_p$. 

\subsection{Diffusion coefficient and Grammage}
\label{sec:grammage}

The calculations presented in this paper allow us to determine the diffusion coefficient rather than assuming it, although the limitations deriving from assuming isotropic and spatially uniform diffusion should always be kept in mind. These limitations are however also common to widely used propagation models such as GALPROP and DRAGON. 

\begin{figure}
   \centering
   \includegraphics[width=0.45\textwidth]{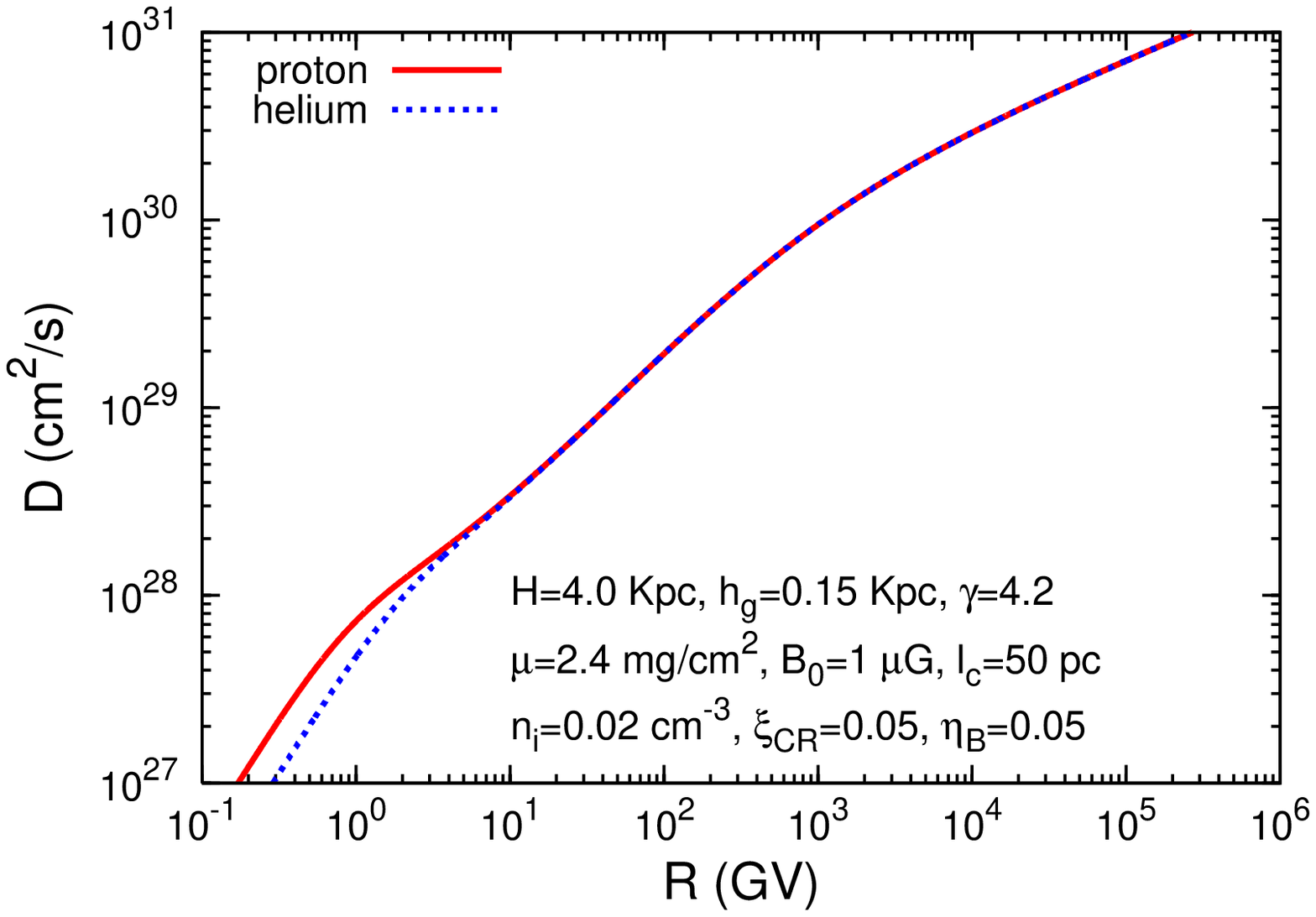}
   \includegraphics[width=0.45\textwidth]{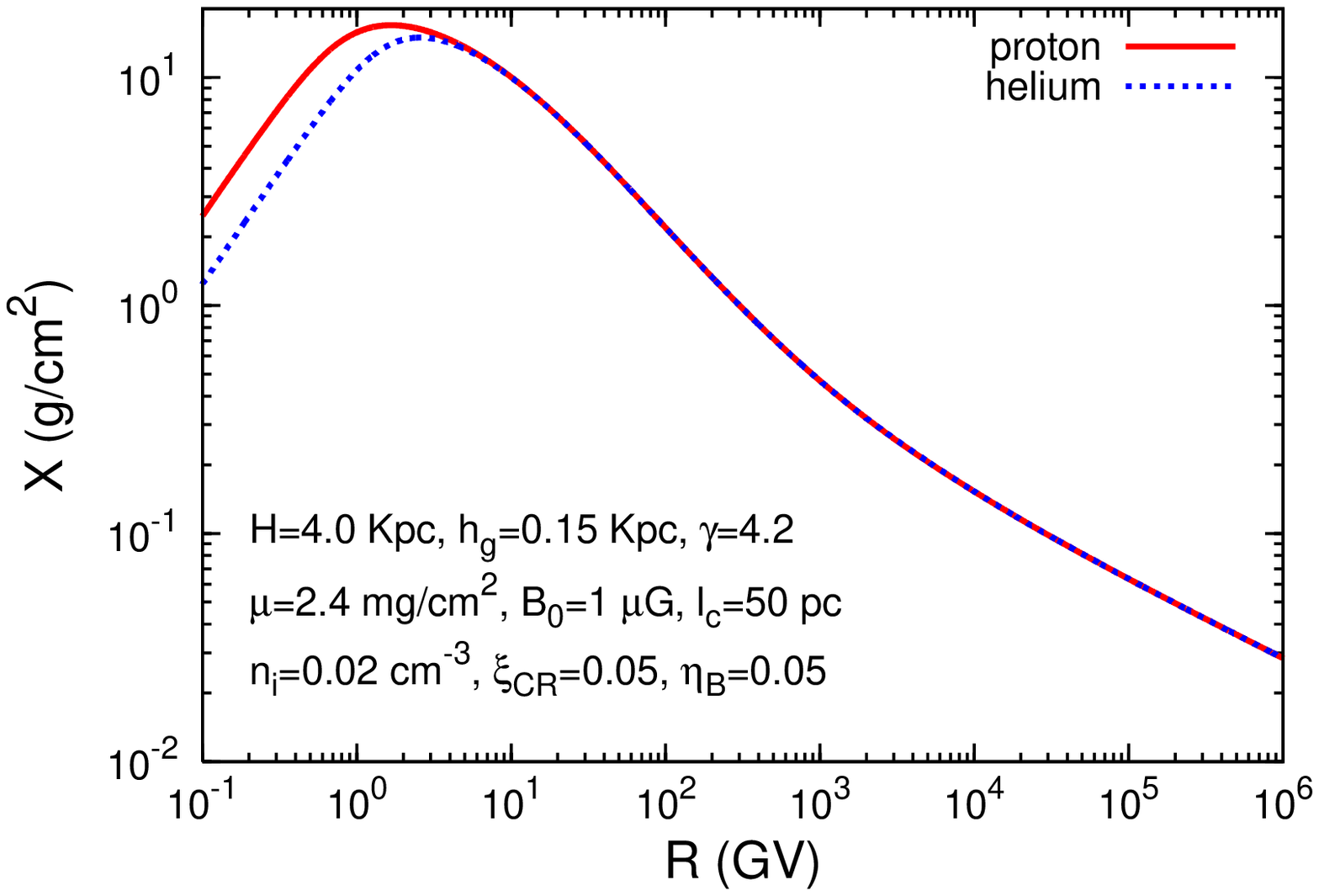}
   \caption{Diffusion coefficient (left panel) and grammage (right panel) as calculated in this paper. The solid (red) line refers to protons, while the dotted (blue) line refers to He nuclei. The diffusion coefficient and grammage for heavier nuclei of given rigidity are indistinguishable from those of He, therefore they are not explicitly shown here.}
   \label{fig:diffusion} 
\end{figure}

The diffusion coefficient plotted as a function of particle rigidity is shown in the left panel of Fig. \ref{fig:diffusion}. The solid (red) line refers to protons while the dashed (blue) line refers to all nuclei. Despite the non-linearity of the problem of self-generation, the diffusion coefficient appropriate for nuclei simply scales with rigidity, therefore we plot here only the diffusion coefficient for protons and helium. Even these two quantities are basically indistinguishable for rigidity $\gtrsim 5$ GV. The change of slope at few hundred GV clearly reflects the transition from self-generation to diffusion in the pre-existing turbulence. The slope of $D_{\alpha}(R)$ at rigidity $10-200$ GV is $\sim 0.6$ (although it is badly approximated by a pure power law). The slope becomes $\sim 1/3$ at rigidity above a few hundred GV. 

The grammage traversed by particles as a function of rigidity is plotted in the right panel of Fig. \ref{fig:diffusion}. The solid (red) line refers to protons, while the dashed (blue) line refers to all other nuclei. The grammage traversed by protons and nuclei at given rigidity is virtually the same, with a small difference only below $\sim 3$ GV. 
It is interesting to note here that the rigidity dependence of the grammage that we obtain at low rigidities ($R\lesssim 1$ GV) is $X\propto \beta=v/c$, that is the same required in propagation models like GALPROP and usually imposed by hand (see for instance \cite{Jones:2001p1956}). However, in the SDM and RAM models, depending on the source injection spectrum, stronger dependences of the grammage on particles velocity have been inferred (see \cite{Ptuskin:2012p2753} and references therein). In our calculations the dependence of grammage on particle velocity $X\propto \beta$ is a natural consequence of the dominance of advection with waves at low rigidities. Above $\sim 10$ GV the grammage steepens to $\sim R^{-0.6}$ while a flattening is found above a few hundred GV. As we discuss below, this trend also reflects in changes of slope in the secondary/primary ratios, such as B/C. 

It is worth recalling that the high energy behaviour ($E\gg TeV/n$) of the diffusion coefficient and grammage is the same as usually predicted by quasi-linear theory for a Kolmogorov spectrum of fluctuations. The value of $\delta B/B_0=\eta_{B}$ needed to explain the high energy fluxes is $\eta_{B}\simeq 0.05$. This is a volume averaged value of $\eta_{B}$ over the whole diffusion region. It is plausible to expect much larger values near the Galactic disc. It is interesting that with this value of $\eta_{B}$, which only depends upon fitting the high energy spectra, and with the standard CR acceleration efficiency ($\xi_{CR}\sim 5\%$) the transition between diffusion in the self-generated turbulence and diffusion in the pre-existing turbulence naturally takes place in the few hundred GV rigidity range \cite{Blasi:2012p2344}.

\subsection{Secondary/Primary and Primary/Primary ratios}
\label{sec:ratios}

In this section we discuss how the physics of self-generation of turbulence by CRs and the effect of pre-existing turbulence reflect on the standard indicators of diffusive propagation, such as the B/C ratio. We also extend the illustration of our results to some primary/primary ratios.

\begin{figure}
   \centering
   \includegraphics[width=1.\textwidth]{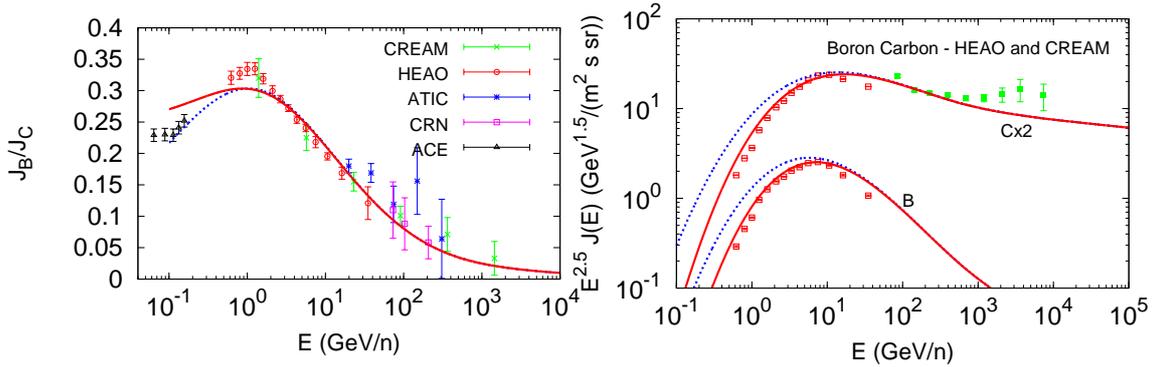}
   \caption{B/C ratio (left panel) and fluxes of B and C nuclei (right panel). The solid (red) lines show the spectra at the Earth, while the dotted (blue) lines show the spectra in the ISM. Data points are from CREAM \cite{Ahn:2009p10593,Ahn:2010p624}, HEAO \cite{Engelmann:1990p10596}, ATIC \cite{Panov:2007p10853}, CRN \cite{Swordy:1990p10799} and ACE \cite{George:2009p10771}.}
   \label{fig:BC} 
\end{figure}

The calculated B/C ratio is compared with data in the left panel of Fig. \ref{fig:BC} while the corresponding boron and carbon fluxes are shown in the right panel. For illustration purposes, in the right panel the C flux has been multiplied by a factor two. The dashed (blue) lines refer to quantities in the ISM, while the continuous (red) lines show the same quantities after accounting for solar modulation. The steep energy dependence of the B/C ratio observed between 1 and 100 GeV/n is well described by our calculations, as a result of the fact that self-generation is very effective in such energy range. 

The quality of the fit to the data is comparable with that obtained by using a standard description of CR propagation (standard diffusion or reacceleration), but here we did not require any artificial breaks in the diffusion coefficient and/or in the injection spectrum. It is also worth stressing that the results illustrated in this paper are not the output of a formal fitting procedure in which the parameter space is scanned systematically and some kind of likelihood method used. Since the calculations are intrinsically non-linear this would be rather time consuming from the computational point of view, therefore we only try to achieve a reasonable fit to the data without actually maximizing a likelihood indicator. At energies around the peak of the B/C ratio, shown in Fig. \ref{fig:BC}, our predicted ratio is somewhat below the observed value. This is not a new result, the same problem at the peak energy $E\lesssim 1$ GeV/n is found also in models with SDM, RAM or diffusion-convection propagation \cite{Ptuskin:2012p2753,Jones:2001p1956}. However it is interesting to notice that  this disagreement is within the spread in measured values that one would see if all data were shown in Fig. \ref{fig:BC} (this can be done using for instance the web interface recently put forward in Ref. \cite{Maurin:2013p11445}).

A better discrimination among different models of the propagation of CRs in the Galaxy could be achieved if accurate measurements of the B/C ratio were available at energies above $\sim 10^{3}$ GeV/n. Some claims of a possible flattening of the B/C ratio are not rare in the literature: if confirmed, such features could be the evidence of the physical phenomenon described in this paper. However, one should keep in mind that spallation reactions inside the sources must be present at some level, so as to reflect in a constant B/C ratio at high energies. At present this so-called ``residual grammage'' (both its normalization and energy dependence) is very uncertain (see \cite{Berezhko:2003p3069,Blasi:2009p135,Mertsch:2009p790}), but it certainly represents a potential contamination for the measurement of the physical phenomenon discussed here. Experimental systematics due to the subtraction of the atmospheric grammage (for balloon-borne experiments) also may hinder the possibility to gather a clean signal of transition of the B/C ratio from a scaling $\propto E^{-0.6}$ to $\propto E^{-1/3}$.

\begin{figure}
   \centering
   \includegraphics[width=0.6\textwidth]{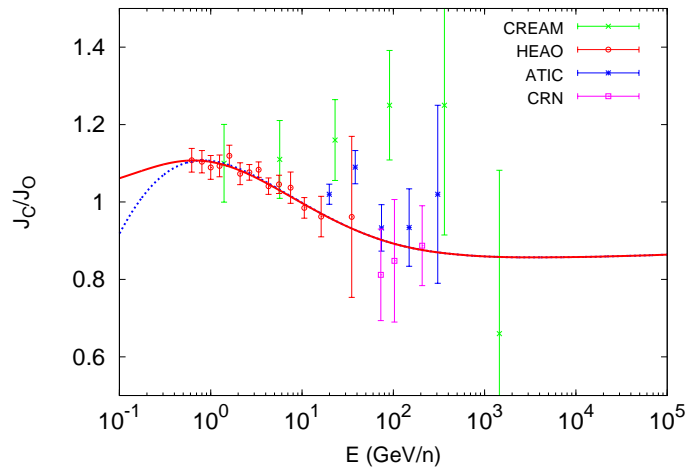}
   \caption{Ratio of Carbon and Oxygen fluxes at Earth (solid red line) and in the ISM (dotted blue line). The small deviation of the ratio from unity and the slight decrease with energy may provide an estimate of the amount of C that is generated as secondary product of spallation of O nuclei. Data points are from CREAM \cite{Ahn:2009p10593,Ahn:2010p624}, HEAO \cite{Engelmann:1990p10596}, ATIC \cite{Panov:2007p10853} and CRN \cite{Swordy:1990p10799}.}
   \label{fig:CO} 
\end{figure}

The ratio of fluxes of primary nuclei can also help constrain the propagation parameters. Carbon and oxygen are both mainly primaries, although a small fraction of carbon can be produced due to spallation of oxygen nuclei. The ratio of C/O is expected to be close to energy independent. The data are not very clear in this respect: the HEAO \cite{Engelmann:1990p10596}, CRN \cite{Swordy:1990p10799} and ATIC \cite{Panov:2007p10853} data suggest that the ratio is roughly constant or even slightly declining with energy, while CREAM data \cite{Ahn:2009p10593,Ahn:2010p624} show a trend to a slight increase of the ratio with energy, though with large error bars. Our predicted C/O ratio is shown in Fig. \ref{fig:CO}, as compared with existing data. A clearly decreasing trend can be seen in the C/O ratio, confirming that a small fraction of C is of secondary origin. 

\begin{figure}
   \centering
   \includegraphics[width=1.\textwidth]{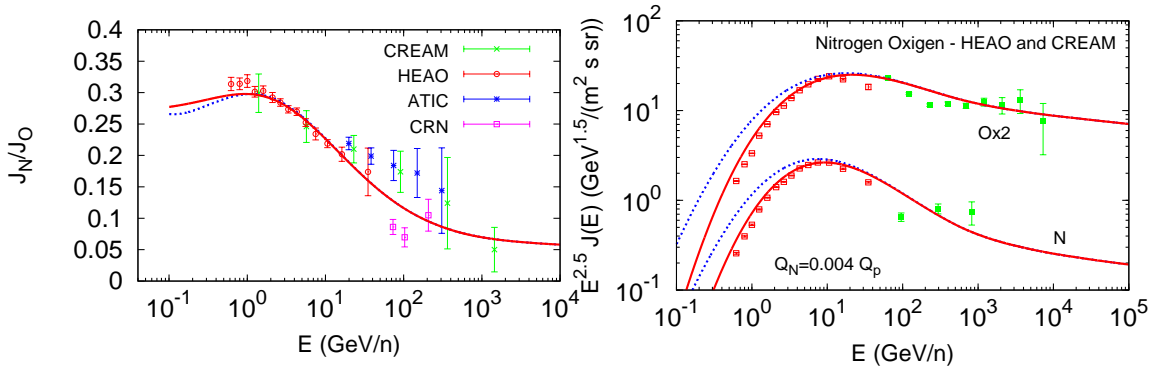}
   \caption{N/O ratio (left panel) and fluxes of N and O nuclei (right panel). The solid (red) lines show the spectra at the Earth, while the dotted (blue) lines show the spectra in the ISM. For the sake of clarity the spectrum of O nuclei in the right panel has been multiplied by a factor two. Data points are from CREAM \cite{Ahn:2009p10593,Ahn:2010p624}, HEAO \cite{Engelmann:1990p10596}, ATIC \cite{Panov:2007p10853} and CRN \cite{Swordy:1990p10799}.}
   \label{fig:NO} 
\end{figure}

An interesting case is that of nitrogen nuclei, which have roughly half primary and half secondary origin. For this reason, one expects that the ratio of fluxes of nitrogen and oxygen shows a more pronounced decline with energy compared with the C/O ratio. Our predicted N/O ratio (left panel) and the fluxes of N and O nuclei (right panel) are shown in Fig. \ref{fig:NO}, as compared with existing data. For illustration purposes the flux of oxygen has been multiplied by a factor two. 

\subsection{Comparison with the SDM and RAM}

The standard calculations of CR transport best explain the data with two classes of models:  1) {\it Standard Diffusion Model} (SDM), in which the injection spectrum is a power law in rigidity and the grammage is a power law $X(R)\propto \beta R^{-\delta}$ with $\delta\simeq 0.6$ for $R>4$ GV and $\delta=0$ for $R<4$ GV. 2) {\it Reacceleration Model} (RAM), in which CR scattering with background turbulence leads to second order Fermi acceleration (typically only important for $R<1-10$ GV) and the injection spectrum is a broken power law $Q(R)\propto R^{-2.4}/\left[1 + (R/2)^{2}\right]^{1/2}$ \cite{Ptuskin:2012p2753,Jones:2001p1956,Seo:1994p2130}. In both classes of models {\it ad hoc} breaks in the energy dependence of either the grammage or the injection spectrum are required in order to fit spectra and secondary to primary ratios at the same time (see for instance \cite{Vladimirov:2012p3043,diBernardo:2010p10666} for recent attempts to achieve such fits). From the theoretical point of view these artificial breaks are not very appealing and most likely they suggest that some important physical ingredients may be missing in the modeling of CR transport, such as the ones discussed in this paper. Clearly, an additional break would be required in these models in order to accomodate the hardening observed by PAMELA and confirmed by CREAM. 

As discussed in \cite{Blasi:2012p2517}, the relatively flat injection spectra predicted by the theory of DSA, very close to $E^{-2}$, would seem to be compatible only with a SDM with $D(E)\sim E^{\delta}$ and $\delta\approx 0.7$. On the other hand, this is known to result in exceedingly large anisotropy at $\gtrsim TeV$ energies, which in fact remains even for $\delta=0.6$ \cite{Blasi:2012p2053}. Taking into account the non-linear effects that stem from the dynamical reaction of CRs on the shock makes this problem even more severe since the injection spectrum becomes harder than $E^{-2}$ at $E\gtrsim 10$ GeV. It has been proposed that the accelerated spectra may be steeper if the velocity of scattering centers is taken into account \cite{Caprioli:2010p133,Ptuskin:2010p1025,Caprioli:2012p2411}, but it is worth keeping in mind that this effect may well make the spectra harder rather than steeper, depending on wave helicity in the shock region.

If the diffusion coefficient is self-generated, as discussed in the present paper, the steep diffusion coefficient at $\gtrsim 200$ GV is due to CRs themselves, and a relatively flat injection spectrum is required $Q(E)\propto E^{-\gamma}$ with $\gamma=2.1-2.2$, that can in principle be accounted for with a mild effect of scattering centers. At energies higher than a few hundred GeV/n, the spectra of individual elements harden so as to make their slope $\sim \gamma+1/3$ if the cascade of waves occurs within the framework of a Kolmogorov cascade. It is quite possible that this scenario may also solve the puzzle of low anisotropy observed at $\gtrsim TeV$ energies, although in order to address this issue one has to take into account the discrete nature of sources \cite{Blasi:2012p2053,Lee:1979p1621,Ptuskin:2006p620}.

\subsection{The case of clouds in the Gould's belt}
\label{sec:clouds}

Two recent papers \cite{Neronov:2012p2217,Kachelrie:2012p10950} have stimulated much discussion since they indirectly confirmed that the spectrum of CRs with energy $10\lesssim E \lesssim 200$ GeV may be steeper than previously thought, and with a slope compatible with the one quoted by PAMELA in the same energy region. The two papers are based on the analysis of the gamma ray emission detected by the Fermi-LAT from selected clouds in the Gould's belt, located appreciably above and below the Galactic disc. The density in the clouds is large enough that the main contribution to the gamma ray emission comes from the generation and decay of neutral pions in inelastic hadronic collisions of CRs with gas in the clouds. The authors of \cite{Neronov:2012p2217} find that the slope of the CR spectrum averaged over all the clouds in the sample is $\sim 1.9$ below $\sim 10$ GeV and $\sim 2.9$ at CR energies $10\lesssim E\lesssim 200$ GeV. The limited Fermi-LAT statistics at high energies does not allow the authors to probe the energy region where, according to PAMELA, there should be an additional spectral break. 

\begin{figure}
   \centering
   \includegraphics[width=.6\textwidth]{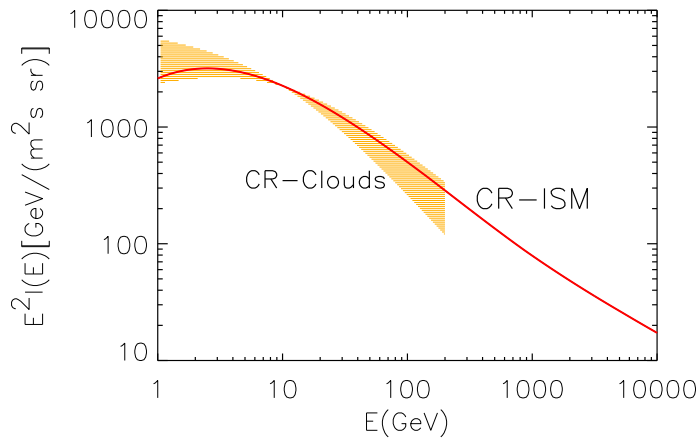}
   \caption{Spectrum of protons in the ISM (thick red line) compared with the spectrum of CRs as inferred from gamma ray observations of clouds in Ref. \cite{Neronov:2012p2217} (shaded area).}
   \label{fig:slope} 
\end{figure}

The low energy behavior of the spectrum inferred by \cite{Neronov:2012p2217} has stimulated much debate since the authors suggest that the effects of solar modulation might be larger than usually thought. This suggestion is mainly motivated by the rather large discrepancy between the CR spectrum inferred from the gamma ray fluxes from clouds and the PAMELA flux of CR protons. It should however be pointed out that the absolute normalization of the CR spectrum in Ref. \cite{Neronov:2012p2217} is quite uncertain since it is based on simple models of the mass distribution in the clouds, and even the mass itself of the clouds is not well known. In this sense the slope of the CR spectrum is better constrained by the analysis of \cite{Neronov:2012p2217} than it is the actual flux of CRs at the clouds' locations.

If confirmed, the CR flux inferred in \cite{Neronov:2012p2217,Kachelrie:2012p10950} could be a measurement of the effect of the ISM CR flux, although we cannot exclude that local effects of stellar modulation near the clouds may be present. On the other hand it is unlikely that such effects would appear in similar ways in different clouds. Clearly the possibility of using gamma ray observations from the Earth to infer the CR spectrum in the ISM is a very well known technique and has been routinely used in the past decade. The real new aspect of the analyses of \cite{Neronov:2012p2217,Kachelrie:2012p10950} lies in the fact that they applied this technique to localized regions of space where the contribution of hadronic CR interactions to the gamma ray production can be considered dominant, due to the high gas density in clouds. Since the CR spectrum is inferred from gamma rays, it is clear that what is being measured is basically the proton spectrum, since protons give the main contribution to pion production. 

In our calculations the CR proton spectrum has a shape that resembles quite closely the one inferred in Ref. \cite{Neronov:2012p2217} (and a similar one derived in \cite{Kachelrie:2012p10950}). The low energy flattening is naturally obtained because of the advection of CRs with the waves that they generate, an effect that should be taken into account in the transport equation even in normal diffusion and reacceleration models, unless the assumption is made that the waves move isotropically with respect to the background magnetic field and the effective mean wave speed vanishes at any location. Waves traveling in all directions are in fact needed in order to have reacceleration. In the case of self-generation induced by CRs, the waves are produced along the CR gradient, therefore there are only forward propagating waves, and the term $v_{A}\frac{\partial f}{\partial z}$ cannot be neglected. The effect of this term is that at low energy the CR spectrum flattens. Self-generation leads to an equilibrium proton spectrum which is rather steep, again in agreement with the observations in \cite{Neronov:2012p2217,Kachelrie:2012p10950}. In Fig. \ref{fig:slope} we show the proton spectrum in the ISM (namely no correction for solar modulation is applied) as resulting from our calculations (solid red line). The shaded area illustrates the shape of the spectrum inferred in \cite{Neronov:2012p2217}, where the normalization has been changed to fit the data at the break energy of 9 GeV/n found in \cite{Neronov:2012p2217}. The area illustrates the error bars on the slope as found in \cite{Neronov:2012p2217}. One can see that the general shape of our predicted proton spectrum is well in agreement with that inferred from gamma ray observations of clouds. Since our prediction also fits the PAMELA data (after correction for solar modulation), we conclude that no anomalous modulation is required, provided one realizes that the cloud mass and mass distribution are very uncertain, which reflects in a fudge factor in the absolute gamma ray fluxes. 

\section{Discussion}
\label{sec:discuss}

We presented a non-linear calculation of the CR transport in the Galaxy in which the diffusion coefficient is calculated as due to the scattering of particles with pre-existing waves and Alfv\'en waves produced by the same CRs due to the excitation of streaming instability \cite{Blasi:2012p2344}. All stable isotopes are included in the calculation. The high energy part of the spectra ($E\gg 1$ TeV/n) is determined almost uniquely by the pre-exisiting turbulence. The low energy part is instead determined by the self-generated turbulence. It is clear that a transition between the two regimes can be expected in an intermediate energy region, as first pointed out in \cite{Blasi:2012p2344}.

The possibility that CRs could be responsible for creating the conditions for their own diffusive motion through excitation of streaming instability was first discussed in Ref. \cite{Skilling:1975p2176,Holmes:1975p621} and summarized in the context of CR confinement in Refs. \cite{Cesarsky:1980p1935,Wentzel:1974p770}. The general conclusion of all these calculations was that either the effect of ion-neutral damping or the NLLD would suppress the waves with sufficiently small wavenumber $k$ faster than they can get excited. Hence CR self-confinement would be possible only for CRs with energy below $\sim 100-200$ GeV/n. The role of ion-neutral damping still requires further investigation: in order for this phenomenon to be neglected one has to assume strong spatial segregation of neutral gas in the Galaxy. This might be reasonable if most neutrals are in molecular clouds, but a small fraction of neutral hydrogen in the disc would be sufficient to damp Alfv\'en waves (both self-generated and pre-existing) very effectively, so as to make CRs free-stream. Diffusion would then be possible only in the halo of the Galaxy \cite{Holmes:1975p621}. We assumed here that segregation is in fact possible in an appreciable fraction of the volume of the Galaxy, so that ion-neutral damping may be neglected.

Our calculations show that a good fit to the proton and helium spectra, to the fluxes of primary nuclei and to the secondary/primary and primary/primary ratios can be obtained by properly taking into account the self-generation, the pre-existing turbulence and the advection with the waves. Quite remarkably, the general features of the proton spectrum as measured by PAMELA are very well reproduced. The spectrum is never really a power law, but it may be locally approximated as a power law. The spectrum is rather flat (slope $\lesssim 2$) at energy below 10 GeV/n. It steepens to a slope $\sim 2.8-2.9$ at $10GeV/n\leq E\leq 200GeV/n$ and it gets harder again (slope $\sim 2.6$) at $E>200$ GeV/n. Throughout our calculations the injection spectrum is just a pure power law in momentum, as predicted by the theory of DSA in supernova remnants. The slope of the injected spectrum is $\sim 2.2$ (if the number of injected particles in the range $dp$ of momentum is normalized as $Q(p)dp$). The growth rate of the instability necessary to explain the data requires that CRs be injected in SNRs with a typical CR acceleration efficiency $\sim 5-10\%$. One should appreciate that within this approach there is no need for artificial breaks in injection spectra and/or diffusion coefficient, contrary to what is usually required by standard calculations of CR transport in the Galaxy, such as GALPROP (see for instance \cite{Vladimirov:2012p3043}).

The diffusion coefficient and the grammage traversed by particles are outputs of the calculations and they are shown to be in excellent agreement with the fluxes of primary nuclei observed at the Earth and with the secondary to primary ratios, such as the B/C ratio. Other observables, such as the C/O and the N/O ratios are also in very good agreement with the data. 

We also compared the predicted proton spectrum in the ISM with the CR spectrum inferred from gamma ray observations of clouds in the Gould's belt \cite{Neronov:2012p2217,Kachelrie:2012p10950}, which is however limited to $E\lesssim 200$ GeV/n. The hardening at $E\lesssim 10$ GeV/n and the soft spectrum (slope $\sim 2.8-2.9$) that we predict both compare well with the results of the analyses in \cite{Neronov:2012p2217,Kachelrie:2012p10950}.

\section*{Acknowledgment}
We are grateful to the anonymous referee for providing several remarks that helped us improve the quality of the paper. We are also grateful to Luca Maccione for several useful conversations on spallation cross sections. We acknowledge continuous scientific collaboration with the rest of the Arcetri High Energy Astrophysics Group E. Amato, R. Bandiera, N. Bucciantini, G. Morlino, O. Petruk and with D. Caprioli and P. Serpico. This work was partially funded through grant PRIN INAF 2010.

\bibliographystyle{JHEP}
\bibliography{crbib}

\end{document}